\documentclass{svjour3}
\usepackage{amsmath}
\def\be{\begin{equation}}
\def\ee{\end{equation}}
\def\bq{\begin{eqnarray}}
\def\eq{\end{eqnarray}}

\journalname{General Relativity and Gravitation}

\begin{document}

\title{Geodesic models generated by Lie symmetries}

\author{ G. Z. Abebe \and S. D. Maharaj  \and  K. S. Govinder}

\institute{
G. Z. Abebe \and S. D. Maharaj  \and  K. S. Govinder \at
Astrophysics and Cosmology Research Unit,
 School of Mathematics, Statistics and Computer  Science,
 University of KwaZulu-Natal, Private Bag X54001, Durban 4000, South Africa\\ \\
\\ \\
\email{maharaj@ukzn.ac.za} \\ \\
}
\date{Received: date / Accepted: date}

\maketitle

\begin{abstract}
We study the junction condition relating the pressure to  the heat flux at the boundary of  a shearing  and expanding  spherically symmetric radiating star when the fluid particles are travelling in geodesic motion.  The Lie symmetry generators that leave the junction condition invariant are identified  and  the optimal system is generated. We use each element of the optimal system  to transform  the partial differential equation to an ordinary differential equation. New   exact solutions, which are group invariant under the action of Lie point infinitesimal symmetries, are found. We obtain families of traveling wave solutions and self-similar solutions, amongst others. The gravitational potentials are given in terms of elementary functions, and the line elements can be given explicitly in all cases. We show that the Friedmann dust model is regained as a special case,  and we can connect our results to earlier investigations.

\keywords{radiating stars \and junction conditions \and Lie symmetries }

\end{abstract}

\section{Introduction}
 Exact models of relativistic radiating stars are required to investigate  physical phenomena such as particle production, temperature profiles, the cosmic censorship hypothesis,  and gravitational collapse of stars. The interior spacetime of a relativistic radiating star  matches with the exterior Vaidya \cite{1} solution.  The junction condition relating  the pressure with the heat flux at the boundary of the star, which was first developed by Santos \cite{11},  must be satisfied.  Kolassis  et al. \cite{2} discussed  the dissipative effects of fluid particles travelling in geodesic motion; the  Friedmann dust solution is regained in the absence of heat flow. The  exact solution  obtained in this model has been widely used to  investigate  physical features of stars. The physical investigations include  modelling  radiating gravitational collapse in spherical geometry with neutrino flux  by Grammenos and Kolassis \cite{3}. Tomimura and Nunes \cite{4} describe realistic astrophysical processes  in the presence of heat flow. The temperature in casual thermodynamics for particles travelling in geodesic motion produces higher central values than the Eckart theory as shown by Govender  et al.\cite{5}.  Thirukkanesh  and Maharaj \cite{100} generated new  classes of solutions for geodesic fluid trajectories. Later Govender and Thirukkanesh \cite{101} extended the model to include a nonvanishing cosmological constant. Recently, Ivanov \cite{102} performed a general  analysis of perfect fluid spheres with heat flow which contains the geodesic model as a special case. Note that these treatments that have been mentioned do not include shear.

  The effects of shear and anisotropic pressure were studied by Chan \cite{6}, Herrera and Santos \cite{7}, Herrera et al.  \cite{8} and Thirukkanesh et al. \cite{800}. Euclidean stars in general relativity may be modelled with nonvanishing shear; in Euclidean stars both the areal and proper radii are equal. Particular shearing solutions were found by Herrera and Santos \cite{6aaaa}, Govender et al. \cite{6ccc} and Govinder and Govender \cite{16aaa}. Naidu et al. \cite{9} obtained the first exact solution  with shear by considering the geodesic motion of fluid particles. Rajah and Maharaj \cite{10} extended this result and obtained new classes of solutions by transforming the junction condition to a Riccati  equation and  solving it.  Thirukkanesh  and Maharaj \cite{10a}  obtained new classes of exact solutions in terms of elementary functions without assuming a separable form for the gravitational potentials. The presence of shear changes the nature of the boundary condition and it is more difficult to integrate in general.

Generating exact solutions to the boundary condition of a shearing radiating star in geodesic motion, using the symmetry approach, is  the main objective of this paper. In the past the Lie theory of differential equations has been  used to great effect in solving the Einstein field equations in cosmology 
\cite{a17,b17,c17,d17,e17,f17,h17} and in higher dimensions \cite{g17}. With the help of the  Lie symmetry theory of differential equations Govinder and Govender \cite{16aaa} obtained an exact solution for the boundary condition for an Euclidean star. We believe that their treatment   has been  the first  group theoretic approach to solve the boundary condition.  Later Abebe et al. \cite{i17}  generated two classes  of  exact solutions  for a conformally flat model by solving the junction condition exactly in the presence of anisotropic pressures using  Lie symmetries. We expect that the application of the Lie symmetry analysis to  a  shearing and expanding radiating star when the fluid particles are in geodesic motion is likely to provide new insights.
 
We briefly introduce the shearing and expanding  model,  when the fluid particles are in geodesic motion, and present the junction condition in Sect. \ref{sec11}.  We use a geometric approach to generate exact solutions. In Sect. \ref{sec22} we  find the Lie point symmetries that are admitted by the  boundary condition.  The optimal system for the symmetries is found to which  all group invariant solutions can be transformed. The junction condition  is transformed into an ordinary differential equation for each symmetry in Sect. \ref{sec1}, Sect. \ref{sec2}  and  Sect. \ref{sec3}. By analysing  the relevant ordinary differential equations, solutions are  found to the boundary condition. We show the connection to known results and obtain limiting metrics.  We make some concluding remarks in Sect. \ref{sec4}.
 
\section{The model \label{sec11}} 
 The  assumption that the fluid particles of radiating stars are in geodesic  motion is reasonable   in relativistic astrophysics modelling.  The line element for the interior spacetime of a radiating star in  geodesic motion with shear and expansion can be  written as
\begin{equation} \label{a1}
ds^2=-dt^2+B^2dr^2+Y^2d\Omega^2 
\end{equation}
where $Y=Y(r,t)$ and $B=B(r,t)$ are the metric functions and $d\Omega^2 \equiv d\theta^2+\sin ^2\theta d\phi^2 $.  The fluid four-velocity \textbf{u} is comoving and is given by
 $$u^a=\delta^a_0$$ 
 and the energy momentum has the form
 \begin{equation}\label{7aa}
 T_{ab}=\left(\mu+p \right) u_au_b+pg_{ab}+q_au_b+q_bu_a+\pi_{ab}
 \end{equation}
where $\mu$ is the energy density of the fluid, $p$ is the isotropic pressure, $q_a$ is the heat flux and $\pi_{ab}$ is the stress tensor.  The heat flow  vector \textbf{q} takes the form 
 $$q^a=(0,q,0,0)$$  
 since $q^au_a=0$ and the heat is assumed to flow in the radial direction. The stress tensor has the form
$$\pi_{ab}=\left( p_\parallel-p_\perp\right)\left(n_an_b-\frac{1}{2} h_{ab}\right)  $$
where $p_\parallel$ is radial pressure, $p_\perp$ is tangential pressure, \textbf{n} is a unit radial vector given by $n^a=\frac{1}{B}\delta^a_1,$ and $h_{ab} $ is the projection tensor. The isotropic pressure is given by
 $$p=\frac{1}{3}\left( p_\parallel+2p_\perp\right) $$
in terms of the   radial pressure, $p_\parallel$, and  the tangential pressure, $p_\perp$.

The kinematical and dynamical quantities can be generated from the general treatment \cite{800} by setting $A=1$. The four-acceleration $\dot{u}^a$, the expansion scalar $\Theta$, and the magnitude of the shear scalar $\sigma$ are given by
\begin{subequations}
 \begin{eqnarray} 
 \dot{u}^a&=&0\\
 \Theta &=&2\frac{Y_t}{Y}+\frac{B_t}{B} \\
 \sigma &=&\frac{1}{3}\left(\frac{Y_t}{Y}-\frac{B_t}{B} \right) 
\end{eqnarray}
 \end{subequations}
respectively  for the line element \eqref{a1}.
 The Einstein field equations for the interior matter distribution become
\begin{subequations}\label{144}
 \begin{eqnarray}
 \mu &=& 2\frac{B_t}{B}\frac{Y_t}{Y}+\frac{1}{Y^2} +\frac{Y_t^2}{Y^2}-\frac{1}{B^2}\left(2 \frac{Y_{rr}}{Y}+\frac{Y_r^2}{Y^2}-2\frac{B_r}{B}\frac{Y_r}{Y}\right)   \\
 p_\parallel&=& -2\frac{Y_{tt}}{Y}-\frac{Y_t^2}{Y^2}+\frac{1}{B^2} \frac{Y_r^2}{Y^2}-\frac{1}{Y^2}\label{144p}\\
 p_\perp&=&\frac{1}{B^2}\left( -\frac{B_r}{B}\frac{Y_r}{Y}+\frac{Y_{rr}}{Y}\right)-\left( \frac{B_{tt}}{B}+\frac{B_t}{B}\frac{Y_t}{Y}+\frac{Y_{tt}}{Y}\right)  \label{14r}\\
 q&=&-\frac{2}{B^2}\left(-\frac{Y_{rt}}{Y} +\frac{B_t}{B}\frac{Y_r}{Y}\right)\label{144d} 
 \end{eqnarray}
 \end{subequations}
for the line element \eqref{a1} and matter distribution \eqref{7aa}. The subscripts stand for partial derivatives with respect to the independent variables $t$ and $r$. The equations \eqref{144} describe the  gravitational interactions in the  interior of a geodesic  shearing and expanding  spherically symmetric star with heat flux and anisotropic pressure. 

The exterior spacetime, describing the region outside the stellar boundary, is described by the Vaidya metric
\be\label{4a1}
ds^2=-\left(1-\frac{2m(v)}{R} \right)dv^2-2dvdR+R^2 d\Omega^2
\ee
The matching of the exterior spacetime  \eqref{4a1} with the interior spacetime \eqref{a1} leads to the following set of junction conditions for the radiating star with shear in geodesic motion
\begin{subequations}\label{rev11}
\begin{eqnarray}
dt&=& \left(1-\frac{2m}{R_\Sigma}+2\frac{dR_\Sigma}{dv} \right)^\frac{1}{2}dv \\
\left(Y \right)_{\Sigma}&=&R_{\Sigma }(v)\\
m(v)&=&\left[\frac{Y}{2}\left(1 +Y_t^2-\frac{Y_r^2}{B}\right)  \right] _{\Sigma} \\
(p_\parallel)_{\Sigma}&=&(Bq)_{\Sigma}\label{san1}
\end{eqnarray}
\end{subequations}
where  $\Sigma $ is the hypersurface that defines the boundary of the radiating sphere.

The junction condition \eqref{san1}  was established by Santos \cite{11} for the first time in the case of shear-free spacetimes. Later it was extended by Glass \cite{11a} for spacetimes with nonzero shear. From \eqref{144p}, \eqref{144d} and \eqref{san1}  we have  
\begin{equation}\label{17}
 2B^2YY_{tt}+B^2Y^2_t-Y^2_r+2BYY_{rt}-2B_tYY_r +B^2=0
\end{equation}
 at the boundary. Equation \eqref{17} is the  fundamental differential equation  governing the evolution of a shearing relativistic radiating star  in geodesic motion.  It is a complicated nonlinear partial differential equation and  difficult to solve in general. We will attempt to integrate  \eqref{17}  to complete the model using the Lie theory of extended groups applied to differential equations.

\section{Lie symmetry analysis\label{sec22}}
As mentioned earlier, the Lie symmetry method has been successfully used to generate exact solutions in General Relativity. In this paper we attempt to find  solutions for the boundary condition of an expanding and shearing radiating star when the fluid particles are travelling in geodesic motion using the Lie symmetry approach.

An $n$th order differential equation 
\be 
F\left(r,t,B,Y,B_r,Y_r,B_{t},Y_t,B_{rr},Y_{rr},B_{rt},Y_{rt},B_{tt},Y_{tt}, \dots\right) =0
\ee
where $B=B(r,t)$ and $Y=Y(r,t)$, admits a Lie point symmetry of the form
\bq 
G&=& \xi_1\left(r,t,B,Y \right)\frac{\partial}{\partial r}+\xi_2\left(r,t,B,Y \right)\frac{\partial}{\partial t}  \nonumber \\
 && +\eta_1\left(r,t,B,Y \right)\frac{\partial}{\partial B}+\eta_2\left(r,t,B,Y \right)\frac{\partial}{\partial Y} \label{sym10}
 \eq   
provided that
\be
\left. G^{[n]}F\right|_{F=0}=0 
\ee
where $G^{[n]}$ is the $n$th prolongation of the symmetry $G$ \cite{12,13}.
The process is  algorithmic and  so can be implemented by computer algebraic packages. 
Using \texttt{PROGRAM LIE} \cite{14}, we find that the junction condition   \eqref{17} admits the infinite-dimensional set of symmetries
\begin{subequations}\label{18}
\begin{eqnarray}
G_1&=&\frac{\partial}{\partial t}\label{1cc}\\
G_2&=&-Bf'(r)\frac{\partial}{\partial B}+f(r)\frac{\partial}{\partial r}\label{3cc}\\
G_3&=&B\frac{\partial}{\partial B}+Y\frac{\partial}{\partial Y}+t\frac{\partial}{\partial t}\label{2cc}
\end{eqnarray}
\end{subequations}
where $f(r)$ is an arbitrary function of $r$. These symmetries tell us that \eqref{17} admits translational invariance (in $t$) as well as scaling invariance in $B$ and $r$ (together) as well as $B$, $Y$ and $t$ (together). When $f(r)$ is constant we have translational invariance in $r$ as well. The appearance of translational invariance in both independent variables indicates that traveling wave solutions may exist.

Each symmetry and any linear combination of the symmetries  may  be helpful in solving \eqref{17}. Any group invariant solution obtained by using these symmetries can be transformed to the group invariant solution obtained by the symmetries in the optimal system which is a subalgebra of the symmetries in \eqref{18}. To obtain  the   subalgebra of the symmetries we begin with the  nonzero vector
\begin{equation}\label{20}
G = a_1 G_1 + a_2 G_2 + a_3 G_3
\end{equation}
 removing   the coefficients $a_i$ of $G$ in a systematic way   using applications of adjoint maps to $G$ to obtain
\begin{subequations}\label{mar7}
\begin{eqnarray}
G_1&=&\frac{\partial}{\partial t}\\
aG_1+G_2&=&a\frac{\partial}{\partial t}-Bf'(r)\frac{\partial}{\partial B}+f(r)\frac{\partial}{\partial r}\\
aG_2+G_3&=&(1-af'(r))B\frac{\partial}{\partial B}+Y\frac{\partial}{\partial Y}+t\frac{\partial}{\partial t}+af(r)\frac{\partial}{\partial r}
\end{eqnarray}
\end{subequations}
which is the optimal set of one-dimensional subalgebras of the symmetries in \eqref{18}.
\section{Invariance under $G_1$\label{sec1}}
Using  the generator 
 \be 
 G_1=\frac{\partial}{\partial t}
 \ee
   we determine the invariants from the surface condition \begin{equation}
\frac{dt}{1}=\frac{dr}{0}=\frac{dB}{0}=\frac{dY}{0}
\end{equation}
We obtain the invariants   $r$ and
\begin{eqnarray}\label{fff}
B=h(r),\quad Y=g(r)
\end{eqnarray}
 for the generator $G_1$. Thus the gravitational potentials are static for the generator $G_1$ and the heat flux \eqref{144d} must vanish. We do not pursue this case further as the star is not radiating. 
\section{Invariance under $aG_1+G_2$\label{sec2}}
The generator
\be\label{varia4}
aG_1+G_2= a\frac{\partial}{\partial t}-Bf'(r)\frac{\partial}{\partial B}+f(r)\frac{\partial}{\partial r}
\ee
 yields the surface condition
\begin{equation}
\frac{dt}{a}=\frac{dr}{f(r)}=\frac{dY}{0}=-\frac{dB}{Bf'(r)}
\end{equation}
from which we determine the  invariants as
\begin{subequations}\label{inv1}
\begin{eqnarray}
x&=&  \int\frac{dr}{f(r)}-\frac{t}{a} \quad\equiv\quad \bar{f}(r)-\frac{t}{a} \label{varia1}\\
B&=&\frac{h\left(x\right)}{f(r)}\\
Y&=&g\left(x\right)\label{varia2}
\end{eqnarray}
\end{subequations}
for the generator $aG_1+G_2$. When $f(r)=1$, our independent variable becomes
\be x = r - \frac{1}{a} t \ee
and so we are in the realm of traveling waves solutions, with wave speed equal to ${1}/{a}$.

Using  the transformation \eqref{inv1} equation \eqref{17} becomes
\begin{equation}\label{3int}
 2 a g'g h'-2 a g  g''h+ \left(a^2+(2 g g''+g'^2)\right)h^2=a^2 g'^2 
\end{equation}
which is a Riccati equation in $h$. It is  difficult to integrate \eqref{3int} in general. We can find particular solutions  for $h$  by specifying the functional form of $g$.

We note solutions to the boundary condition with dependence on the variable $x$ given in \eqref{varia1} have not been found previously for particles traveling in geodesic motion in the stellar interior. The geodesic model of  Thirukkanesh  and Maharaj \cite{10a}, containing earlier models, has the potential
\bq
Y&=&  \left[ R_1(r)t+R_2(r) \right] ^{\alpha}\label{varia3}
\eq
where $\alpha=2/3$, $ 1$. The functional dependence in \eqref{varia2} is different from that in \eqref{varia3} since $g(x)$ is arbitrary. It is only in special cases, for particular forms of $R_1(r)$ and $R_2(r)$, that \eqref{varia2} can be brought into the form \eqref{varia3}. Therefore the classes of solution corresponding to \eqref{inv1} are new. This is not surprising since the  Thirukkanesh  and Maharaj \cite{10a} models were generated using  a method that transforms \eqref{17} to a first order separable equation. In our case we are seeking group invariant solutions to the second order differential equation \eqref{17} using the Lie theory of differential equations. We achieve this by restricting the coefficients in \eqref{3int} which generate forms for the function $g$.

\subsection{ Case I: $2 g g''+g'^2=0$\label{sss}}
 We set 
\be \label{ayzo1}
 2 g g''+g'^2=0
\ee
 Then \eqref{ayzo1} can be integrated to give
\be\label{xx1}
g(x)=(b+c x)^{2/3}
\ee
where $b$ and $c$ are arbitrary constants of integration.
On substituting \eqref{xx1} into \eqref{3int}, we have
 \be \label{xx2} 
12 c (b+c x)h'+4 c^2 h+9 a (b+c x)^{2/3} h^2=4 a c^2
\ee
which is a simpler Riccati equation in $h$. On integration we obtain
\be \label{xax3}
h(x)=\frac{2c }{3 (b+c x)^{1/3}}\left( \frac{\exp\left[ \frac{3 a}{ c}  (b+c x)^{1/3}+d\right]-1 }{\exp \left[ \frac{3 a}{ c}  (b+c x)^{1/3}+d\right]+1 } \right) 
\ee
where $d$ is an arbitrary constant of integration. 

The gravitational functions are given by
\begin{subequations}\label{ss1}
\begin{eqnarray}
B&=&\frac{2}{3}\frac{c }{ f(r)\left(b+c \left(\bar{f}(r) -\frac{t}{a}\right)\right)^{1/3}} \nonumber \\
&& \times \left( \frac{\exp\left[ \frac{3 a}{ c}  \left(b+c \left(\bar{f}(r) -\frac{t}{a}\right)\right)^{1/3}+d\right]-1 }{\exp \left[ \frac{3 a}{ c}  \left(b+c \left(\bar{f}(r) -\frac{t}{a}\right)\right)^{1/3}+d\right]+1 }\right) 
\\
&&\nonumber\\
Y&=&\left( b+c\left( \bar{f}(r)-\frac{t}{a} \right) \right) ^{2/3}
\end{eqnarray}
\end{subequations}
which satisfy the boundary condition \eqref{17}. The line element becomes 
 \begin{eqnarray} 
ds^2&=&-dt^2+\frac{4}{9}\left[\frac{c }{ f(r)\left(b+c \left(\bar{f}(r) -\frac{t}{a}\right)\right)^{1/3}} \right. \nonumber \\
&& \left. \times \left( \frac{\exp\left[ \frac{3 a}{ c}  \left(b+c \left(\bar{f}(r) -\frac{t}{a}\right)\right)^{1/3}+d\right]-1 }{\exp \left[ \frac{3 a}{ c}  \left(b+c \left(\bar{f}(r) -\frac{t}{a}\right)\right)^{1/3}+d\right]+1 }\right)  \right] ^2dr^2\nonumber\\
&&+ \left( b+c\left( \bar{f}(r)-\frac{t}{a} \right) \right) ^{4/3}d\Omega^2 
\end{eqnarray}
 We believe that this is a new solution for geodesic motion. 

The functional dependence on the spacetime variables  $t$ and $r$ in the metric functions is that of  a travelling wave. This becomes clearer if we select  a particular form of the function $f(r)$. We consider the special case $f(r)=1$, $a=-1$, $b=0$, $c=1$, $d=0$.  Then we obtain the line element  
\bq
ds^2 &=&-dt^2+\frac{4}{9}\left[\frac{1}{ \left(r+t\right)^{1/3}}\left( \frac{\exp\left( 3\left( r+t \right)^{1/3}\right) -1 }{ \exp\left( 3 \left( r+t \right)^{1/3}\right) +1 }\right) \right] ^2dr^2  \nonumber \\
&& +\left[ r+t\right]^{4/3}d\Omega^2   \label{thiru2}
\eq
 which has an explicit  travelling wave solution form. (Observe that the special case \eqref{thiru2} arises essentially  since the generator  \eqref{varia4} has  the reduced form $aG_1+G_2= -\frac{\partial}{\partial t}+\frac{\partial}{\partial r}$.)  The spacetime is well-behaved as translational invariance under  $t$ removes the singularity that would occur at $t=r=0$.

\subsection{ Case II: $2 a g  g''= 0$}
 If we set 
 \be 
 2 a g  g''= 0
 \ee
  then a simple integration  gives
\be \label{nnn1}
g(x)=bx+c
\ee 
Equation \eqref{3int} becomes
\be\label{linea3}
2 a b(c+b x) h'+\left(a^2+b^2\right) h^2=a^2 b^2
\ee
which is also a  Riccati equation in $h$. This can be integrated to give
\be \label{nnns1}
h(x)=\frac{ab}{\sqrt{a^2+b^2}} \left( \frac{ d \left( c+ b x\right) ^{\frac{\sqrt{a^2+b^2}}{ b}}-1}{ d  \left( c+b x\right) ^{\frac{\sqrt{a^2+b^2}}{ b}}+1}\right) 
\ee
where $d$ is a nonzero constant. 

The gravitational potentials have the form
\begin{subequations}\label{ss42s}
\begin{eqnarray}
B&=&\frac{a b}{f(r)\sqrt{a^2+b^2} } \left( \frac{   d\left(c+b\left(\bar{f}(r) -\frac{t}{a}\right)\right) ^{\frac{\sqrt{a^2+b^2}}{ b}}-1}{d \left( c+b \left(\bar{f}(r) -\frac{t}{a}\right)\right) ^{\frac{\sqrt{a^2+b^2}}{ b}}+1}\right) 
\nonumber\\
&&\\
Y&=&c + b \left(\bar{f}(r) -\frac{t}{a}\right)
\end{eqnarray}
\end{subequations}
which is a solution for the boundary condition \eqref{17}. The line element becomes
 \begin{eqnarray} \label{lineelt}
ds^2&=&-dt^2+\left[\frac{a b}{f(r)\sqrt{a^2+b^2} } \left( \frac{ d \left( c+ b \left(\bar{f}(r) -\frac{t}{a}\right)\right) ^{\frac{\sqrt{a^2+b^2}}{ b}}-1}{d  \left(c+b \left(\bar{f}(r) -\frac{t}{a}\right)\right) ^{\frac{\sqrt{a^2+b^2}}{ b}}+1}\right)\right] ^2 dr^2\nonumber\\
&&+\left[c + b \left(\bar{f}(r) -\frac{t}{a}\right)\right]^2d\Omega^2 
\end{eqnarray}
 We believe that this solution is new. 
 
 A simple form can be regained from \eqref{lineelt}. We set $a=-1$, $b=1$, $d=1$, $c=0$ and  $f(r)=1 $ to obtain
\begin{eqnarray} 
ds^2=-dt^2+\frac{1}{2}\left[\left( \frac{  \left[ \left(r +t\right)\right] ^{\sqrt{2}}-1}{  \left[ \left(r +t\right)\right] ^{\sqrt{2}}+1}\right)\right] ^2 dr^2+\left[ \left(r +t\right)\right]^2d\Omega^2 
\end{eqnarray}
  in terms of elementary functions. The time translational symmetry ensures that this spacetime is also well-behaved. Again this particular metric has the corresponding reduced Lie generator $aG_1+G_2= -\frac{\partial}{\partial t}+\frac{\partial}{\partial r}$ as in Sect. \ref{sss}.

\subsection{Case III: $a^2+2 g g''+g'^2=0$}
We  have also considered  the case  
 \be \label{varia5}
 a^2+2 g g''+g'^2=0
 \ee
 The condition  \eqref{varia5} can be integrated  to give
  \begin{equation}\label{c3}
\pm\frac{b}{a^3} \arctan\left( a\sqrt{\frac{g(x)}{b-a^2g(x)}}\right) \mp\frac{\sqrt{bg(x)-a^2g(x)^2}}{a^2} =x+c
\end{equation} 
This an implicit solution involving the function $g(x)$. For this case \eqref{3int}
becomes
\be \label{para12}
2  g'g h'-2  g  g''h=a g'^2 
\ee
which is a linear equation in $h$. This equation may be integrated to obtain
\be \label{nnn}
h(x)=g'\left( \frac{a}{2 }\int \frac{dx}{ g}  +d\right)
\ee
where $d$ is a constant of integration,  and \eqref{nnn} gives $h$ in terms of $g$ explicitly.
The gravitational potentials therefore are
 \begin{subequations}\label{ss42sa}
\begin{eqnarray}
B&=& g'\left( \frac{a}{2 }\int \frac{dx}{ g}  +d\right) \\
Y&=&g
\end{eqnarray}
\end{subequations} 
 which are functions of $x=\bar{f}(r)-\frac{t}{a}$, and $g$ is given by \eqref{c3}. The line element is 
\begin{eqnarray} \label{para}
ds^2&=&-dt^2+\left[g'\left( \frac{a}{2 }\int \frac{dx}{ g}  +d\right)\right] ^2 dr^2+g^2d\Omega^2 
\end{eqnarray}
 which is written in terms of the function $g$ only.
 
 It is possible to rewrite the solution given above parametrically. We introduce a new parameter $u$ for convenience. Then \eqref{c3} can be represented as 
\begin{subequations}\label{para1}
\bq
x&=&\frac{b}{a^3}\left(u-\frac{1}{2}\sin(2u) \right) -c\label{para32}\\
g&=&\frac{b}{2a^2}\left( 1-\cos(2u)\right) 
\eq
\end{subequations}
in a parametric representation. In this case the line element \eqref{para} becomes
\bq
 ds^2 &=&-dt^2+\left[\frac{b}{2a^2}\left( 1-\cos(2u)\right) \right] ^2 \nonumber \\
 && \times \left\lbrace \left[\frac{b }{a^2}\left(d-\frac{a^2}{b}\cot (x) \right)  \sin(2x)  \right]^2 dr^2+d\Omega^2 \right\rbrace 
\eq
where the relationship between the $u$ and $x$ is given by \eqref{para32}.

\section{Invariance under $aG_2+G_3$ \label{sec3}}
By using the generator
\be 
aG_2+G_3= (1-af'(r))B\frac{\partial}{\partial B}+Y\frac{\partial}{\partial Y}+t\frac{\partial}{\partial t}+af(r)\frac{\partial}{\partial r}
\ee
 we find the invariants from the invariant surface condition
\begin{equation}
\frac{dt}{t}=\frac{dr}{af(r)}=\frac{dB}{(1-af'(r))B}=\frac{dY}{Y}
\end{equation}
which are given by
\begin{subequations}\label{mar5}
\begin{eqnarray}
x&=& \frac{t}{\exp\left( \int\frac{dr}{af(r)}\right) }\quad\equiv \quad\frac{t}{\exp\left(\tilde{f}(r)\right) } \\
B&=&h(x)\frac{\exp\left( \int\frac{dr}{af(r)}\right) }{f(r)}\quad= \quad h(x)\frac{\exp\left(\tilde{f}(r)\right)}{f(r)}\\
Y&=&g(x)t
\end{eqnarray}
\end{subequations}
for the symmetry $aG_2+G_3$. We observe that the new independent variable $x$ has a self-similar form.  This is particularly evident when $f(r)=r/a$ as then we have 
\begin{equation}
x = \frac{t}{r}
\end{equation}

With  transformation \eqref{mar5} equation \eqref{17} becomes
\bq
&& 2 a x^3 g'g  h'- 2 a x^2 g \left(2 g'+x g''\right) h \nonumber \\
&& +a^2 \left(1+g^2+x^2 g'^2+2 x g \left(3 g'+x g''\right)\right) h^2=x^4 g'\qquad  \label{mar2}
\eq
Equation \eqref{mar2} is a different Riccati equation in $h$ from those considered previously. As in Sect. \ref{sec2} we can find group invariant solutions  to \eqref{mar2} by restricting the coefficients which produce functional forms for the function $g$.

\subsection{Case I: $g^2+x^2 g'^2+2 x g \left(3 g'+x g''\right)=0$}

In this case we have
\be 
 g^2+x^2 g'^2+2 x g \left(3 g'+x g''\right)=0 \label{new1}
 \ee
 To integrate we set
\be p(x) = (xg(x))^{3/2}   \label{new2} \ee
Then
\eqref{new1} 
becomes
\be p''(x) = 0 \ee
with solution
\be p(x) = b + c x \ee
 and so \eqref{new1} has the general solution
\be\label{op3}
 g(x)=\frac{(b+c x)^{2/3} }{x}
\ee 
where $b$ and $c$ are arbitrary constants of integration.
The transformation \eqref{new2}  was picked out due to the fact that \eqref{new1} possesses eight Lie point symmetries and so is linearisable to the free particle equation.

On substituting equation \eqref{op3} into \eqref{mar2} we have
\bq
&& 6 a \left(3 b^2+4 b c x+c^2 x^2\right)h'-4 a c^2 x h-9 a^2 (b+c x)^{2/3} h^2 \nonumber \\
&& =-\left(9 b^2+6 b c x+c^2 x^2\right) \label{vvv}
  \eq
which is also a Riccati equation in $h$. This can be integrated to give
 \begin{eqnarray}\label{same} 
h(x)&=&\frac{(3 b+c x)}{3 a (b+cx)^{1/3}}\left(\frac{1-k+\left( k+1\right) \exp\left(\frac{3}{ c} (b+c x)^{1/3}\right) }{1-k-\left( k+1\right) \exp\left(\frac{3 }{ c}(b+c x)^{1/3}\right) }\right)
\end{eqnarray}
  where $k$ is an arbitrary constant of integration. 

The gravitational potentials become
 \begin{subequations}\label{sol1001}
\bq
B&=&\frac{\exp\left( \tilde{f}(r)\right) \left(3 b+\frac{c t}{\exp\left( \tilde{f}(r)\right) }\right)}{3 af(r) \left(b+\frac{c t}{\exp\left( \tilde{f}(r)\right) }\right)^{1/3}}
 \nonumber \\
 && \times \left( \frac{1-k+\left( k+1\right) \exp\left(\frac{3}{ c} \left(b+\frac{c t}{\exp\left( \tilde{f}(r)\right)   }\right)^{1/3} \right) }{1-k-\left( k+1\right) \exp\left(\frac{3}{ c} \left(b+\frac{c t}{\exp\left( \tilde{f}(r)\right) }\right)^{1/3} \right) }\right)  \nonumber\\
&&\\
Y&=&\exp\left( \tilde{f}(r)\right) \left(b+\frac{c t}{\exp\left( \tilde{f}(r)\right) }\right)^{2/3}
\eq
\end{subequations}
which is a particular solution for the master equation.  The line element is 
\bq
ds^2&=&-dt^2+\left[\frac{\exp\left( \tilde{f}(r)\right) \left(3 b+\frac{c t}{\exp\left( \tilde{f}(r)\right) }\right)}{3 a f(r) \left(b+\frac{c t}{\exp\left( \tilde{f}(r)\right) }\right)^{1/3}}
 \right. \nonumber \\
&& \times 
\left. \left( \frac{1-k+\left( k+1\right) \exp\left(\frac{3}{ c} \left(b+\frac{c t}{\exp\left( \tilde{f}(r)\right)   }\right)^{1/3} \right) }{1-k-\left( k+1\right) \exp\left(\frac{3}{ c} \left(b+\frac{c t}{\exp\left( \tilde{f}(r)\right) }\right)^{1/3} \right) }\right)   \right] ^2 dr^2\nonumber\\
&&+\left[\exp\left( \tilde{f}(r)\right) \left(b+\frac{c t}{\exp\left( \tilde{f}(r)\right) }\right)^{2/3}\right]^2d\Omega^2  \label{newmet1} 
\eq
We believe that the metric \eqref{newmet1} has not been found before. Note that the gravitational potentials are given in terms of elementary functions. There is simplification in the form of the line elements for particular values of the parameter $k$. We consider these cases below.
\subsubsection{ $k=\pm1$}
 If we set the arbitrary constant of integration $k=\pm1$ in equation \eqref{newmet1} then we obtain the metric
\begin{eqnarray} 
ds^2&=&-dt^2+\left[\frac{\exp\left( \tilde{f}(r)\right)  \left(3b+\frac{c t}{\exp\left( \tilde{f}(r)\right) }\right)}{3a f(r)  \left(b+\frac{c t}{\exp\left( \tilde{f}(r)\right)  }\right)^{1/3} }\right] ^2 dr^2  \nonumber \\
&& + \exp\left( 2\tilde{f}(r)\right) \left(b+\frac{c t}{\exp\left( \tilde{f}(r)\right)   }\right)^{4/3}  d\Omega^2 
 \label{fre1} 
 \end{eqnarray}
which is a simple form.

If we set $  \exp\left( \frac{1}{3}\tilde{f}(r)\right)   =r$, $c=1$ and $b=0$ in \eqref{fre1} then we have
\begin{equation} \label{b2}
ds^2= -dt^2+t^{4/3} \left[  dr^2+r^2d\Omega^2 \right] 
\end{equation}
which is the  Friedmann dust model in the absence of heat flux. This is a desirable feature as the Friedmann dust model also arises as a special case in the analyses of Kolasis et al. \cite{2}, Thirukkanesh and Maharaj \cite{100,10a}, Naidu et al.  \cite{9} and Rajah and Maharaj \cite{10}.

 \subsubsection{$k=0$}
 The case  $k=0$ in equation \eqref{newmet1} is also of physical interest. This case contains earlier investigations of geodesic configurations with shear experiencing gravitational collapse. Hence the group invariant approach followed in this paper does regain other physically viable models.
If we set $ \exp\left(\frac{1}{3} \tilde{f}(r)\right)  =\beta(r)$, $b=0$, $c=1$  and
 utilize  a  translation of time ($t \rightarrow t+\alpha $)   then \eqref{newmet1} becomes
\begin{equation} \label{b1}
ds^2=-dt^2+(t+\alpha)^{4/3}\left[\beta'(r)^2  \left( \frac{1+\exp\left(   \frac{3(t+\alpha)^{1/3}}{\beta(r) }\right) }{1-\exp \left( \frac{3(t+\alpha)^{1/3}}{\beta(r) }\right)}\right)^2dr^2+\beta(r)^2d\Omega^2 \right] 
\end{equation}
 This   solution  is related to the first category of the Rajah and Maharaj \cite{10} models.  In the Rajah and Maharaj \cite{10} model there is an additional function of integration which is absent in \eqref{b1}.  When this quantity is set to be unity  then \eqref {b1} is exactly the same as the Rajah-Maharaj metric. Observe that if we  further set $\beta(r) =r$ and $\alpha = 0$ then 
 the line element \eqref{b1}  has the form
\begin{equation}\label{b12} 
ds^2=-dt^2+t^{4/3}\left[  \left( \frac{1+\exp\left(   \frac{3t^{1/3}}{r }\right) }{1-\exp \left( \frac{3t^{1/3}}{r }\right)}\right)^2dr^2+r^2d\Omega^2 \right] 
\end{equation}
This solution was  first  obtained by  Naidu et al. \cite{9} for a shearing radiating star in geodesic motion. (Note that the arbitrary function of integration in the Rajah-Maharaj metric has to be set to be unity in \cite{9} to regain \eqref{b12}. Thus, in the solution space of \eqref{17}, their set of solutions overlap with our set of solutions.)  They analysed heat dissipation and  pressure anisotropy   and showed that this was a realistic description of matter configuration undergoing gravitational collapse. 

  For sufficiently large values of $t$  the expression $ \left( \frac{1+\exp\left(   \frac{3t^{1/3}}{r }\right) }{1-\exp \left( \frac{3t^{1/3}}{r }\right)}\right)^2$ approaches  unity  and the line element \eqref{b12} is approximately
\begin{equation} \label{b22}
ds^2\approx -dt^2+t^{4/3} \left[  dr^2+r^2d\Omega^2 \right] 
\end{equation}
which is  the limiting  Friedmann dust model in the absence of heat flux. In the Rajah and Maharaj \cite{10} model the Friedmann dust model is regained exactly when a function of integration is set to be zero. In the case of the metric \eqref{b12}  the dust model only arises approximately.

If we set $ \exp\left( \tilde{f}(r)\right)  =r$, $b=0$, $k=0$ and $c=1$  in \eqref{newmet1} then we have
\begin{equation} 
ds^2= -dt^2+\left( \frac{t}{r}\right)^{4/3} \left[\frac{1}{9}\left( \frac{1+\exp\left[ 3 \left( \frac{t}{r }\right)^{1/3}\right] }{1-\exp\left[ 3  \left( \frac{t}{r}\right)^{1/3}\right]}\right)^2  dr^2+r^2d\Omega^2 \right] 
\end{equation}
 in terms  of the self-similar variable $t/r$.  We observe that  the self-similar variable indicates  the existence of a homothetic Killing vector.  Wagh  and Govinder \cite{17}
 found a self-similar  vector in  shearing spherically symmetric spacetimes. Recently, Abebe et al. \cite{i17} obtained new models for  a conformally flat radiating star in which a particular class contains the self-similar variable.

\subsection{Case II:  $ 2 g'+x g''=0$} 
If we set 
\be 
 2 g'+x g''=0
 \ee
  then we obtain the function 
\be\label{x1}
g(x)=\frac{b}{x}+c
\ee
where $b$ and $c$ are arbitrary constants of integration. On  substituting equation \eqref{x1} into \eqref{mar2} we have
\be \label{ricca1}
 2 ab (b+c x)  h'- a^2 \left(1+c^2 \right) h^2=-b^2 
\ee
which is a Riccati equation in $h$. This  can be integrated to give
\be\label{x2}
h(x)=\frac{b }{a \sqrt{1+c^2}}\left(\frac{1-k \left(b+ c x\right)^{ \frac{\sqrt{1+c^2}}{c}}}{1+k\left( b+ c x\right)^{ \frac{\sqrt{1+c^2}}{c}}} \right) 
\ee
where $k $ is a nonzero constant.

 The gravitational potentials have the form
\begin{subequations}\label{x3}
\begin{eqnarray}
B&=&\frac{b \exp\left( \tilde{f}(r)\right) }{ a\sqrt{1+c^2}f(r) }\left(\frac{1- k\left(b+\frac{c  t}{\exp\left( \tilde{f}(r)\right)}\right)^{ \frac{\sqrt{1+c^2}}{c}}}{1+k\left(b+\frac{ c  t}{\exp\left( \tilde{f}(r)\right)}\right)^{ \frac{\sqrt{1+c^2}}{c}}} \right) \\
Y&=& b \exp\left( \tilde{f}(r)\right)+c t
\end{eqnarray}
\end{subequations}
 which is a particular solution for the master equation \eqref{17}. The metric is given by
 \begin{eqnarray} \label{lin1}
ds^2&=&-dt^2+\left[\frac{b \exp\left( \tilde{f}(r)\right) }{a \sqrt{1+c^2}f(r) }\left(\frac{1-k\left(b+\frac{ c  t}{\exp\left( \tilde{f}(r)\right)}\right)^{ \frac{\sqrt{1+c^2}}{c}}}{1+k\left(b+\frac{c  t}{\exp\left( \tilde{f}(r)\right)}\right)^{ \frac{\sqrt{1+c^2}}{c}}} \right) \right] ^2 dr^2\nonumber\\
&&+\left[b \exp\left( \tilde{f}(r)\right)+c t\right]^2d\Omega^2 
\end{eqnarray}
which is another group invariant model.

 If we set  $\exp\left( \tilde{f}(r)\right)=r$, $a=1/\sqrt{2}$, $b=1$, $c=1$ and  $k=1$  then  the line element \eqref{lin1} becomes
\begin{eqnarray} \label{thiru3}
ds^2&=&-dt^2+\left( \frac{1-\left(1+\frac{ t}{ r}\right)^{\sqrt{2}}}{ 1+\left(1+\frac{  t}{r}\right)^{\sqrt{2}} }\right) ^2dr^2+r^2 \left(1+\frac{t}{r}\right)^{2}d\Omega^2 
\end{eqnarray}
which has a simple form.

\section{Conclusion \label{sec4}}
We  considered a shearing and expanding relativistic  radiating star when the fluid particles are in geodesic motion. This model was analysed with the Lie infinitesimal generators applicable to differential equations. We studied in particular the junction condition  which relates the radial pressure to the heat flux. Three Lie point symmetries admitted by this equation were found   and an optimal system was  obtained. The symmetries were  used to reduce the governing highly nonlinear partial differential equation  to  ordinary differential equations. By solving the reduced ordinary differential equations, and transforming to the original variables, we obtained  exact   solutions for the master equation. It was particularly pleasing to observe that we were able to provide families of traveling wave solutions as well as families of self-similar solutions.  Both types of solutions have be found to have great application in a variety of areas of mathematical physics \cite{tw,ss}.

 Our classes of solutions contain new  and  previously obtained solutions. (We utilised the computer software package Mathematica \cite{180} for some of the integrations and to verify the correctness of all solutions.) We regained the Friedmann dust model as a special case of one family of solutions. In addition, the connection to the previous models of Naidu \emph{et al.} \cite{9} and  Rajah and Maharaj \cite{10}  was shown. Therefore we have demonstrated  that the Lie method is a useful tool in modelling gravitational behaviour in collapse. 
 
 This approach has allowed us to solve the rather complicated equation \eqref{17}.  In its original form, this equation is very difficult to solve.  We used appropriate group invariants to reduce the equation to Riccati ODEs. While this allowed us to make some progress, the resulting equations did not yield to the standard approaches for solving Riccati equations.  However, by making simplifying assumptions, we {\it were} able to solve the equations.  It is remarkable that the simplifications also ensured that the standard techniques could be applied.  In fact, the simplified Riccati equations could then be transformed into second order linear equations with constant coefficients!  This is a elegant happenstance and completely unexpected. None of this would have been revealed if not for the Lie symmetry approach.

 The physical features of the models generated here will be studied in greater detail in the future. In particular, the traveling wave collapse and self-similar collapse should be of great interest.

\begin{acknowledgements}
GZA and KSG thank the  University of KwaZulu-Natal and National Research Foundation  for continuing support.
SDM acknowledges that this work is based upon research supported by the South African Research Chair Initiative of the Department of Science and Technology and the National Research Foundation.\end{acknowledgements}

\end{document}